\documentclass[letterpaper, 10pt, conference]{ieeeconf}
\IEEEoverridecommandlockouts
\usepackage{cite}
\usepackage{amsmath,amssymb,amsfonts}
\usepackage{algorithmic}
\usepackage{graphicx}
\usepackage{textcomp}
\usepackage{xcolor}
\usepackage{verbatim}
\usepackage{url}
\usepackage{subcaption} 
\usepackage{adjustbox}
\usepackage{array}
 \usepackage{hyperref}
 
\def\BibTeX{{\rm B\kern-.05em{\sc i\kern-.025em b}\kern-.08em
    T\kern-.1667em\lower.7ex\hbox{E}\kern-.125emX}}
\begin{document}

\title{\LARGE \bf
    RoboBuddy in the Classroom: Exploring LLM-Powered Social Robots for Storytelling in Learning and Integration Activities
}


\author{ Daniel Tozadore, Nur Ertug, Yasmine Chaker and Mortadha Abderrahim 
\thanks{All the author are affiliated to the Computer-Human Interaction in Learning and Instruction (CHILI) Lab, EPFL, Lausanne, Switzerland. Daniel Tozadore is also affiliated to the Depart of Computer Science of the University College London, London, UK.
Email: \href{mailto:d.tozadore@ucl.ac.uk}{d.tozadore@ucl.ac.uk}. 
}}


\maketitle

\begin{abstract}
%
Creating and improvising scenarios for content approaching is an enriching technique in education. However, it comes with a significant increase in the time spent on its planning, which intensifies when using complex technologies, such as social robots. Furthermore, addressing multicultural integration is commonly embedded in regular activities due to the already tight curriculum. Addressing these issues with a single solution, we implemented an intuitive interface that allows teachers to create scenario-based activities from their regular curriculum using LLMs and social robots. We co-designed different frameworks of activities with 4 teachers and deployed it in a study with 27 students for 1 week. Beyond validating the system's efficacy, our findings highlight the positive impact of integration policies perceived by the children and demonstrate the importance of scenario-based activities in students' enjoyment, observed to be significantly higher when applying storytelling. Additionally, several implications of using LLMs and social robots in long-term classroom activities are discussed.


\end{abstract}


\section{Introduction}

Technology is constantly challenging the way teachers and students interact in primary schools. On the one hand, students have access to interactive devices earlier in life than previous generations, and their exposure to such applications has changed their attention span capacity. On the other hand, while teachers are willing and curious to use new technologies in their classes, they are already overwhelmed with their workload. The complexity of dealing with new technologies in terms of time allocation and cognitive load often weighs against their decision to adopt tools or methodologies with which they are not familiar~\cite{siyami2023investigating}.

Social Robots are strong representatives of this phenomenon. They remain an uncommon device in classrooms as a regular tool, with significantly more complex setup requirements compared to mobile devices and computers already used by teachers. Yet, their usage can extend students' engagement, motivation, and even their learning gains compared to traditional or other digital methods~\cite{woo2021use}. Most often, their advantage relies on the tangible Scenario-Based Learning (SBL) activities that their hardware, combined with their social skill simulation, affords~\cite{shipton2023scenario}.

\begin{figure}
    \centering
    \includegraphics[width=0.95\linewidth]{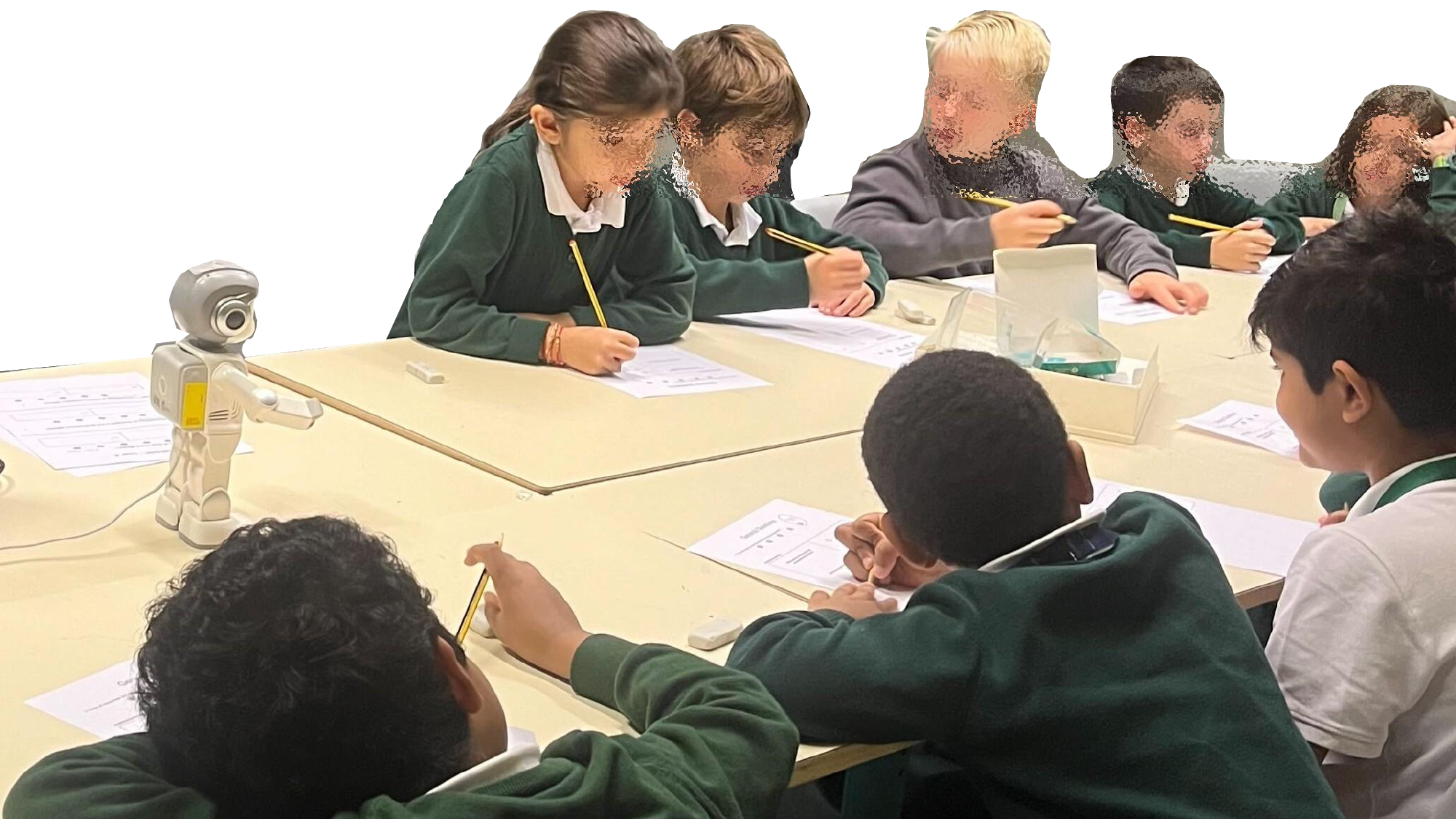}
    \caption{Students engaged in activities with social robots.}
    \label{fig:week}
\end{figure}

Scenario-Based Learning, such as imagination games, role-playing games, and storytelling, are powerful strategies in pedagogical activities and can get potentialised by social robots engaging young students, as exemplified in Fig.~\ref{fig:week}. Students under these conditions tend to be more involved and engaged in the narrative and, therefore, in the content. Furthermore, the simulative nature of these scenarios provides a safe environment for children to exercise their thinking and problem-solving skills, training their attitudes for real-life situations, while remaining playful~\cite{elgarf2024fostering}.
Nonetheless, the diversity of scenario-based activities to maintain novelty for students heavily relies on teachers' creativity, which, as previously mentioned, can be suffocated by their other duties.

The advent of Large Language Models (LLMs) suggested shedding light on this problem, especially when feeding autonomous dialogue and potential behaviour of the robot in Human-Robot Interaction (HRI) activities. So far, some works have addressed the topic of deploying LLM-powered social robots in educational setups.
For instance, using social robots for explanations before Makerspace classes suggested an increase in some students' attention and performance during the following practical session \cite{tozadore2024teachers}. Teachers adapted their explanations to storytelling mode without adding any extra time to their preparations.
Similarly, researchers used social robots powered by LLMs to investigate young students' creativity and learning performance. Their findings suggested that children performed better and were more creative when the robot also displayed more creative thinking \cite{elgarf2024fostering}. 
Although teachers' opinions about social robots have been widely investigated in the last decades, their opinions on LLMs, especially when powering social robots, are still underexplored. Furthermore, opinions before and after using these systems may vary, reinforcing the need for real-world and long-term assessment. A validation of how LLM can address multicultural diversity in classrooms, as well as teachers' opinions on this, is also a crescent concern. 

Therefore, the contribution of this work is three-fold:
(1) proposing an intuitive interface for teachers to generate Scenario-Based Learning (SBL) activities and questions and answers to be immediately deployed to social robots;
(2) a fine-tuning process for a local and lightweight LLM specialized in educational content creation; and
(3) a 1-week experiment in a school, using the system in different educational activities.
Four frameworks of activities were designed with teachers and performed with their young students from different nationalities:
(i) storytelling with questions posed by the robot,
(ii) questions created by the students,
(iii) teachers controlling the prompts and the robot,
(iv) the robot teaching words in children's primary languages. A  feedback session with students and teachers was performed at the end.
To the best of our knowledge, this is the longest study in classrooms of LLM-powered social robots  using regular curriculum content, with teachers participating in the activities.

All the material developed in this project, including the source code of the interface and the LLMs, the content used to prompot the activities, detailed results, and extra material are accessible in the project's GitHub repository\footnote{\url{https://github.com/dtozadore/MI2US_Year2s.git}}.

\section{Related Work}
\label{sec:related_work}



Mathemyths pioneered the integration of mathematical language into narratives through co-creative storytelling, demonstrating how prompt engineering can optimize LLMs to achieve educational outcomes comparable to human-guided storytelling \cite{Zhang2024-qj}. Subsequent research, such as Storypark, further expanded this interactive approach, enabling children to contribute through guided questioning and sketch-based interactions, leading to improved engagement and comprehension of story concepts \cite{ye2024connectionneedmultimodalhumanai}.  
StoryBuddy further advances human-AI collaboration by offering flexible parental involvement and automatic question generation, enabling personalized educational goals and progress tracking \cite{Zhang_2022}. Complementing these approaches, a Question-Answer Pair Generation system proposed in ``It is AI's Turn to Ask Humans a Question" enhances reading comprehension by generating targeted QA pairs from children's storybooks, underscoring the critical role of education-specific models and datasets in improving pedagogical performance~\cite{yao2022aisturnaskhumans}.

Specifically with social robots, the use of agents to stimulate children's creativity, showing notable advantages of autonomous models designed for storytelling is being already investigated~\cite{elgarf2024fostering}. Outcomes revealed enhanced creativity in children—especially in fluency, flexibility, and elaboration—when interacting with a creative autonomous robot, emphasizing the importance of tailored robot behaviour for maximizing learning gains. 
%
Related to the tutors opinions about LLM, instructors of maker-space classes used storytelling provided by LLM to storify the theoretical explanations of the sessions \cite{tozadore2024teachers}. Outcomes showed its high potential for generating multilingual, context-appropriate stories and suggested that adding social behaviours, like encouragement and sentiment analysis, could further enhance student engagement and expectation management. 
However, these works do not provide analyses about LLM's  flexibility when used to create different types of activities and operated by teachers.


\section{The System}\label{sec:methods}


To achieve our goal of affording LLM-powered social robots for SBL activities, we implemented three distinct modules based on their functionality. First, an interactive web interface was developed for teachers to program storytelling using standard content, which can be adapted for scenario-based activities. 
Secondly, a web application was created to facilitate communication between the interface and LLM modules. Initial testing was conducted using ChatGPT 3.5 Turbo, which was enhanced through prompt engineering techniques and subsequently applied to a locally trained LLM. Finally, a fine-tuned LLM was implemented to run on a laptop without requiring an internet connection, ensuring that data remains locally processed and stored.

\subsection{Teachers' interface and Prompt engineering}

\begin{figure} [h] 
    \centering
    \includegraphics[width=1\linewidth]{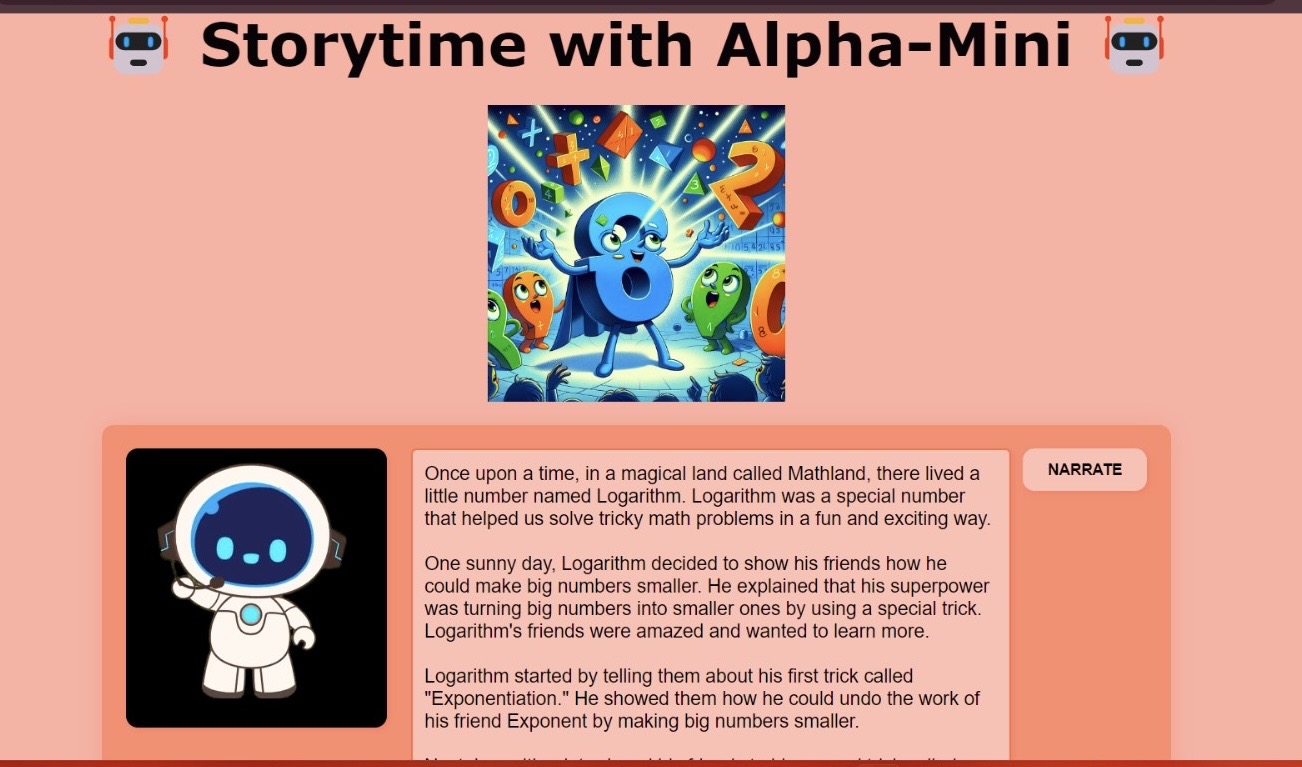}
    \caption{System interface screenshot.}
    \label{fig:interface}
\end{figure}



To facilitate teachers' interaction with the system, we designed an interface that runs on any internet browser, as shown in Fig.~\ref{fig:interface}. Through this interface, teachers (or experimenters) can prepare a variety of activities and select from a pool of predefined settings, which are then translated into specialized prompts to feed the LLM. The frontend of the interface was developed using HTML and CSS, while the backend was implemented in Python 3.8, utilizing Flask to enable communication between the frontend and backend.
We based our design in feedback given by 10 instructors of maker-space activities that participated in studies with the previous version of the interface~\cite{tozadore2024teachers}. 



The initial design of the interface supported the creation of activities in two modes:
\textit{Story Generation} and \textit{Lecture Storification}.
Later, a third mode, \textit{Robot Lecture Explanation}, was introduced for use in a baseline condition, as further detailed in Section~\ref{sec:experiment}.


After the teacher sets the language preference, various levels of AI assistance and content generation options are offered to enhance the story generation process and facilitate the creation of lectures.
For the \textit{Story Generation} process, five levels of AI assistance are provided, ranging from 0 (no help) to 4 (maximum AI assistance).
For the \textit{Lecture Storification} setting, three levels of AI assistance are available, starting at 0 (complete lecture content provided for generation) and going up to 2 (maximum AI assistance, where only the lecture topic is provided to the AI).
Finally, the \textit{Robot Lecture Explanation} mode generates content explanations based on the characteristics of the role that the agent was prompted to assume.


 This section is followed by the age group specification phase, which is crucial for tailoring storytelling to the developmental needs of the target audience. An age-based setting approach tries to ensure that the content of the story aligns with the child’s comprehension abilities, avoiding material that is either too simple or too complex. Similarly, the age-appropriate approach for storytelling targets fostering social and emotional development, since classifying stories based on the specific needs of each age group tend to lead to optimal outcomes in terms of information retention and cognitive and emotional development.


%

The age appropriation approaching is tailored through few-shot prompting approach, starting with simple and repetitive language for toddlers to build vocabulary. It then progresses to relatable narratives for preschoolers that foster logical reasoning and empathy. Stories for early elementary students introduce moral lessons and problem-solving, while late elementary stories emphasize adventure-driven plots with conflicts and resolutions. For pre-teens, the model focuses on critical thinking and ethical dilemmas. 

In designing the system prompts for each generation model, both zero-shot and few-shot prompting techniques were utilized. Zero-shot prompting involves directly instructing the model to perform a task without providing any examples or demonstrations. Conversely, few-shot prompting leverages in-context learning by including demonstrations in the prompt to guide the model towards better performance. These demonstrations condition the model for subsequent tasks, improving its response generation.
For the Q\&A session generation and regeneration process, Chain-of-Thought prompting was employed. This technique encourages the model to articulate intermediate reasoning steps, enhancing its ability to produce coherent and logically sound answers.

After story generation is completed, teachers can enhance the story in three ways:
(i) directly editing the story using the keyboard,
(ii) requesting a regeneration of the story based on a new prompt provided by the teacher, or
(iii) asking for a completely different story generated by the AI.

The goal of this process is to grant teachers autonomy, using the AI as a supportive tool rather than the primary element in course creation. Similarly, teachers can insert questions about the generated content or ask the generative model to propose questions. After the teacher confirms which questions to proceed with, the system prompts whether they would like to save the story and transition to the \textit{executing scenario} state.

In the \textit{executing scenario} state, the teacher can choose between virtual robot support or a social robot option to animate the course using the generated story and additional Q\&A sessions facilitated by the robot assistant.

If the virtual mode is selected, an animated robot avatar (as the one shown in  the bottom left part of Fig.~\ref{fig:interface}) is made available. Teachers can instruct the avatar to narrate the story using the \textit{pyttsx3} text-to-speech library\footnote{\url{https://pypi.org/project/pyttsx3/}}. The avatar delivers the generated content in the selected language. Alternatively, if connected to the Alpha Mini robot, the manufacturer's text-to-speech library is used for narration.
To further enrich the scenarios and assist students in answering questions, image generation support is included. The DALL·E-3 model is used for generating images, with the complexity of the images tailored to the age group. For younger age groups, more saturated and visually stimulating images are prioritized to enhance engagement. For older age groups, the focus of the images shifts to the main character of the storyline.




A full demonstration of the system interface can be seen at the demo video on YouTube\footnote{\url{https://www.youtube.com/watch?v=CyGm-Kk3QA8 }}.

\subsection{PhinetunEd: a Phi model fine-tuned in education} \label{ssec:llm}


Initially, the OpenAI Python API was used to test our interface with the GPT-3.5 Turbo version\footnote{\url{https://platform.openai.com/docs/models/gpt-3.5-turbo}}, as it was the best-performing LLM available at the time of interface development. However, due to concerns about data privacy and the need for improved performance in pedagogical tasks, we opted to explore alternative solutions and implement a local model.


Therefore, given our experimental setup's limited computing resources and the importance of timely response generation, we only considered Small Language Models, mainly those with around 3 billion parameters or less. In particular, we consider the Phi-3-Mini-128k-instruct model\cite{Abdin2024-wd}. The model has 3.8B parameters and a 128K context window and was considered the state-of-the-art open model in our targeted LLM size at the time of the fine-tuning. We use QLoRA to fine-tune the model on a consumer GPU~\cite{dettmers2023qlora}. QLoRA leverages quantization and low-rank matrix adapters to compress the parameters to a 4-bit representation and reduce the number of trainable parameters. This technique performs on par with traditional fine-tuning while reducing the computation load and memory footprint.
For our training data, we sampled a subset of 2500 data samples from Cosompedia~\cite{benallal2024cosmopedia}. This synthetic dataset includes diverse educational resources such as textbooks, blog posts, stories, and WikiHow articles for various audiences; in particular, it provides instruction following tasks for story generation covering common sense and day-to-day knowledge. Since our proposed model is a Phi model fine-tuned for education, it was given the name of PhinetunEd.

We evaluated our models on three different tasks, assessing their capabilities in terms of general knowledge, story completion, and story generation. The results of the models on these tasks are reported in Table~\ref{tab:model_performance}. For general knowledge, we considered the MMLU benchmark~\cite{mmlu}. The benchmark covers various subjects at different difficulty levels, including STEM, humanities, and social sciences. We ran the evaluation of both the base model quantized and our fine-tuned model that we call PhinetunEd locally using the \textit{lm-evaluation-harness} framework~\cite{eval-harness}. In contrast, we reported measures from OpenAI's technical report for the GPT model~\cite{openai2024gpt4technicalreport}. For story completion and reasoning, we considered Hellaswag~\cite{zellers2019hellaswagmachinereallyfinish}. In this task, an event description is provided, and the model must select the most appropriate completion from two suggested options. We used the same measurement method as described above. For story generation, we sampled 200 prompts from the WritingPrompts dataset~\cite{writingprompts}, extracted from an online forum where users submit story premises and prompts. Stories were generated using the three models and evaluated using OpenAI's GPT-4-turbo\footnote{\url{https://platform.openai.com/docs/models/gpt-4-turbo}} as a judge~\cite{llm-judge}. The judge is tasked to evaluate the stories on different aspects (grammar, coherence, relevance to the premise, creativity, engagement, and educational value) by attributing a rating on a scale from 1 to 10. 

\begin{table}[h]
\centering
\begin{tabular}{|c|c|c|c|}
\hline
Model & Knowledge & Story Completion & Story Generation \\ 
 & MMLU & Hellaswag & WritingPrompts \\ 
\hline
GPT-3.5-turbo & \textbf{70\%} & \textbf{85.5\%} & 6.87 \\ 
\hline
Phi-3 base & 62\% & 59\% & 7.05 \\ 
\hline
PhinetunEd & 60\% & 58\% & \textbf{7.46} \\ 
\hline
\end{tabular}
\caption{Models' benchmark performances comparison.}
\label{tab:model_performance}
\end{table}

The benchmark evaluation showed that the GPT model performs better in knowledge, reasoning, and story completion, whereas our finetuned models score best in story generation. Moreover, a qualitative analysis over multiple generations revealed that the GPT model was performing slightly better in narratives, with more words focusing on the fantastic aspects of the story (``in the Mythical forest of ...", ``in the enchanted world of ..."). In contrast, the finetuned Phi model provided substantially more depth in the theoretical explanation of the prompted topics embedded in the story. Therefore, we opted for leaving both models available for story generation in our system, with the user deciding which model to use in each interaction. 


\section{Experiment}\label{sec:experiment}

To test our system in real-world application for classrooms, we performed a 1-week experiment using an Alpha Mini robot\footnote{\url{https://www.ubtrobot.com/en/consumer/humanoidRobots/alphaSeries/AlphaMini}} connected to our system. The following exploratory research questions were used to guide this study: 



\begin{itemize}
    \item \textbf{RQ1:} How can storytelling impact scenario-based activities with social robots?   
    \item \textbf{RQ2:} What particular kinds of activities do students prefer when interacting with LLM powered social robots?
    \item \textbf{RQ3:} How do children perceive the embedded elements of social inclusion? 
     \item \textbf{RQ4:} What are teachers' opinions after using LLM-powered social robots for scenario-based activities?

\end{itemize}

To investigate \textbf{RQ1}, one class (Year 2-1) performed the first two activities in the \textit{Lecture to Storytelling} mode, while the second class (Year 2-2) was exposed to the \textit{Robot Lecture Explanation} mode.
To explore \textbf{RQ2}, we co-designed different activities with teachers, exploiting the functionalities of our system.
To analyse \textbf{RQ3}, we intrinsically added integration elements to the personality and background of the robot. 
Examples of used integration elements were the background of the robot, coming from another planet and asking children for help in its integration, and sentences the robot said reinforcing this background, such as, ``\textit{in my observations of your world, I learned that metal is very hard to be twisted, while elastic can be easily twisted. Is it true?}" 
Finally, to investigate \textbf{RQ4}, beyond including the teachers in the initial design, they also participated in the activities and in the evaluation workshop at the end. 

\subsection{Participants and their primary languages}
A total of 27 students from an international school in Switzerland. Their average age was 6.42 (SD 0.49) years old and and they presented 20 different nationalities. The students were enrolled in two different classes: 14 in Year 2-1, where 6 distinct languages were reported, being the primary language: English (n=5), Russian (n=3), Spanish (n=2), Italian (n=2), Romanian (n=1), and Ukrainian (n=1); and 13 students in Year 2-2 that spoke 7 primary languages, consisting of English (n=3), Mandarin (n=3), Turkish (n=2), Spanish (n=2), Japanese (n=1), Marathi (n=1), and Serbian (n=1). Nevertheless, all the students had good to perfect understanding of English, since that was the language used by all the school staff.


 




Four female teachers also participated in this study:
2 leading teachers and 2 auxiliary teachers, one pair for each of the two classes. We consider demographic data of the teachers not relevant to this study, thus they are not reported.
All teachers and guardians of the children signed the permission to use their data in the anonymized form, as approved in the ethical committee for this study\footnote{EPFL HREC process number 073-2023}.

\subsection{Co-designing with the teachers}

 In a kick-off meeting of 50 minutes via Zoom, we collaborated with the leading teachers of the two Years to develop 4 different activities to be performed during the dates of the experiment. In the following week, the sessions were taken every day, where students and teachers of the same class (Year 2-1 and Year 2-2) came to the room where the experiment was set up and participated in the corresponding activity of that day for 50 minutes. Detailed descriptions of the activities are provided in the next subsections.

Moreover, we showed the outcomes from the two available language models GPT-3.5 and PhinetunEd to the teachers and asked for their opinions. Teachers claimed that the stories generated by GPT could be more engaging due to the nature of the used words, while PhinetunEd provided more room for intervention from the teachers to dive deeper into concepts of the topics. These observations were aligned to the ones presented in Section~\ref{ssec:llm}.

\subsection{Resulting activities}
For all the activities, the system was set to the appropriate age and the specific setup for each activity is presented next. 

\subsubsection{\textbf{Monday - The Robot presentation \& Robot asks questions}} The robot presented itself as RoboBuddy, the new peer of the class that would take a 1-week adventure on Earth, and decided to join the students in their classes. The presentation story was generated with GPT-3.5. After that, the robot approached the content of object propriety changing (How can we mold things into shape by bending, twisting, 
stretching and squashing?) according to their condition (Storytelling or Explanation). At the end, the robot asked students questions about the approached content, also generated by the model.
Contrary to the introduction story, both the story and questions were created with PhinetunEd model.


\subsubsection{\textbf{Tuesday - Questions from the children}}

The robot approached the multiple applications of material molding (art, daily use of object, manufacturing, etc), again according to the groups condition. Students were instructed to pay attention and come up with questions to the robot at the end, instead of simply answering questions from the robot. One by one, the students asked their questions where PhinetunEd model was also used for both ends.

\subsubsection{\textbf{Wednesday - Teachers open-style interaction}}

The two first activities were prepared in collaboration with teachers but run by the experimenters. However, in this session, the teachers were the ones performing a free style interaction with the system in real-time. They were allowed to use any resources available that they wanted, like creating stories, using the questions that the robot generated, taking students' questions, and so on.  
Both teachers from Year 2-1 and 2-2 opted for the topic approach for solid object transformations in the mode \textit{Lecture Storifying} and full AI generation using PhinetunEd.

\subsubsection{\textbf{Thursday - Robot learns their languages}}

Students were asked to share a word they understood in their native language or another language they knew. One at a time, children would say the word aloud along with its meaning in English. After all the other children repeated the word, Robo-buddy was challenged to create a short paragraph incorporating the chosen word. The system was designed to generate a sentence in English that used the word's meaning while preserving the original word in the text.
For example, one student chose the word ``correr" (meaning ``to run" in Spanish), and the LLM generated a paragraph for the robot to say: ``Sometimes when I am running late to school, I like to `correr' with my friends. `Correr' is also a hobby that I do with my friends during the lunch break."
Most words in other languages were mispronounced by the robot, as its text-to-speech was set for English. However, we addressed this by explaining that the robot was still learning other languages and therefore struggling with new pronunciations. This workaround was perceived as a factor of empathy by the children towards the robot. GPT was used for this task due to its more generalized language knowledge compared to PhinetunEd, and the experimenters controlled the robot.

\subsubsection{\textbf{Friday - Feedback workshop}}

On the last day of activities, we explained the robot's programming and system functionalities to the students in layman's terms. We provided a brief overview of each performed activity, followed by soliciting students' feedback about the activities. At the end, students were asked to provide their overall perception, rate how much they liked each activity on a 5-point Likert scale, and select their favourite one.

For each day, we designed and implemented pre- and post-evaluation tasks, asking students to draw key characteristics of the content before and after the interventions. However, very few were able to complete the task, and most of those who did simply repeated their initial drawings. As a result, we concluded that assessing learning gains in this study would not be feasible.

\begin{figure*}
    \centering
    \vspace{0.4cm}
    \includegraphics[width=0.99\linewidth]{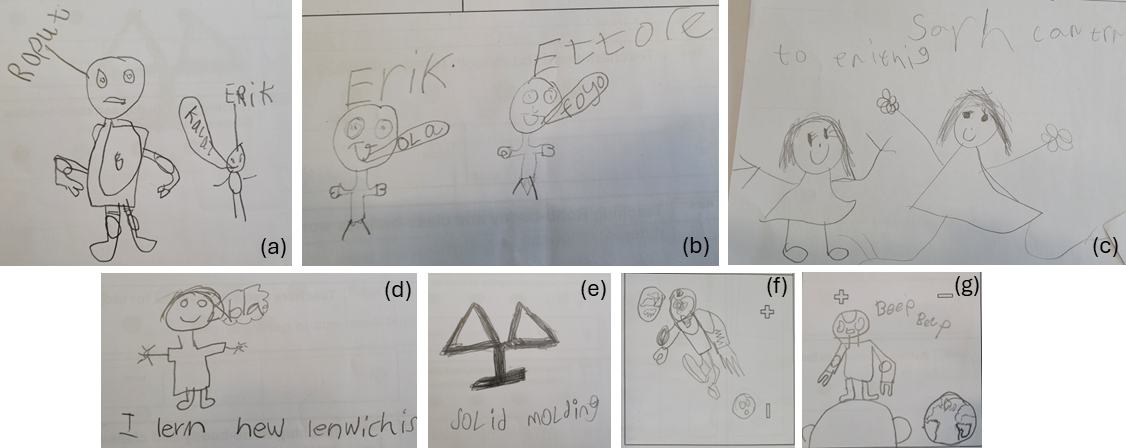}
    \caption{Children drawings in the final day: (a) student speaking with the robot in their mother tongue, (b) students communicating in each other's language, (c) friendship between students, (d) student claiming they learned new languages, (c) content approached in the story, (f) robot sadly returning to its planet, and (g) robot sad alone at its planet.}
    \label{fig:draws}
\end{figure*}

\section{Results \& Discussion}\label{sec:results}


\subsection{Students enjoyment} \label{ssec:students}

\begin{figure} [!ht]
    \centering
    \includegraphics[width=.5\textwidth]{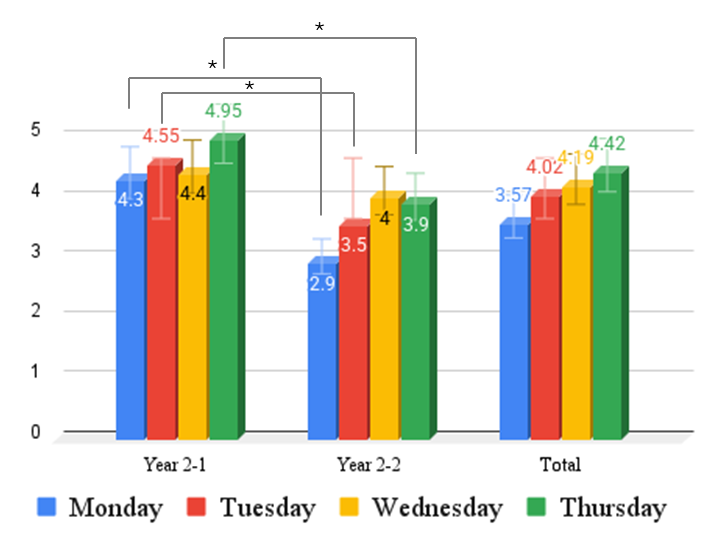}
    \caption{Students enjoyment scores in the activities.}
    \label{fig:likert}
\end{figure}

In the \textit{Feedback Workshop}, students rated how much they liked each activity on a 5-point Likert scale and the results of the scoring are illustrated in Fig. \ref{fig:likert}. 
The average for all the activities given by Year 2-1 students was 4.55 (SD 0.63), and the average given by Year 2-2 was 3.58 (SD 1.04).

A significant difference was found between the enjoyment of activities with storytelling, performed with student from Year 2-1, and with explanation group, Year 2-2, both on Monday (W=25, p-value=0.033) and on Tuesday (W=86, p-value=0.024). Not surprisingly, the day that both groups had the storytelling activity, on Wednesday, no difference was noticed (W=73,	p-value=0.177). Again, a significant difference was found (W=72, p-value=0.04) in the language activity on Thursday.

When asked about their preferred activity, the results were coherent. Most of the students (n=6) from Year 2-1 pointed out a preference for the activity on Thursday, when they were invited to speak in their primary language to the robot. On the other hand, the majority (n=5) of students from Year 2-2 ranked the activity on Wednesday, the teacher on control, as their favourite.
%
%




\subsection{Students feedback and takeaways}

Complementary to the rating, students could voice their opinions about what they liked and disliked during this set of activities. The drawing could be a positive or negative aspect they noticed or anything related to their experience with the robot. 
Nothing different from the already expected ``we liked the robot and would prefer doing it more often" was observed from their vocal opinion. 

However, the drawings brought meaningful outcomes from their perception that cover a wide variation of aspects that our system is capable of triggering. 
Seven out of 13 (53.8\%) students from  Year 2-1 provided a drawing, against only 4 out of 14  (28.5\%) from Year 2-2.
When analysing the drawings, 7 of them were worth bringing to this analysis and are presented in Fig.~\ref{fig:draws}, where the other 4 were simply illustrations of the robot alone.


First, students recognized the robot's potential as a peer capable of communicating in their own language, as illustrated in Fig.~\ref{fig:draws} (a). More importantly, the robot was not the most notable element of the interaction. After ``connecting" the children, the robot disappeared from the spot light, enabling the students to feel more excited and comfortable communicating with their peers, even attempting to speak each other's languages, as demonstrated in Fig.~\ref{fig:draws} (b)~and~(c).

The educational aspects of the activities were also noticed by the children, as observable in the Fig.~\ref{fig:draws} (d) in which the student self-portrays with the saying ``I lern new lenwichis", meaning ``I learn new languages", as well in Fig.~\ref{fig:draws} (e) with a metallic structure written ``solid mold".
Lastly, the drawings in pictures Fig.~\ref{fig:draws} (f) and (g), in which the robot is sadly returning to its planet or already there with a sad expression, suggest that the participants identified the emotional element of integration. This hypothesis can be widely explored to create empathy and provide feasible scenarios for children's disclosures related to the topic of inclusion and well-being.

\subsection{Teachers feedback}

When interviewed separately after the experience with our systems, teachers were enthusiastic about the AI tool's ease of use for generating differentiated lessons, especially for multilingual students. They suggested adding interactive games and progress tracking. Some recommended ensuring age-appropriate language and incorporating stories that promote empathy and address bullying.
%
%
Teachers found the tool highly beneficial for lesson planning, differentiation, and student engagement. The Year 2-1 teacher praised its ease of use and adaptability for multilingual students, calling it a ``game changer" for educators. The Year 2-2 teacher highlighted its potential for both classroom learning and at-home revision, appreciating the engaging storytelling aspect. Support teachers also saw great value in the tool—one noted its usefulness for generating comprehension stories, while the other envisioned its application in complementary lessons, particularly for teaching empathy and addressing bullying through personalized narratives.



\subsection{Discussion}


Regarding the influence of storytelling in the activities, addressed in our RQ1, students exposed to this strategy compared to the expository explanation of the robot presented higher self-reported enjoyment than the other group, not only in the activities of different conditions but also in the one with the language. This fact highlights the importance of presenting scenario-based activities, where the students are presented with an application scenario of the content being approached and have to deal with the questions proposed in that context. 


For RQ2, addressing students' activity preferences, a notable divergence was observed both between classes and among individual students. While most students in Year 2-1 preferred the language activity, students from Year 2-2 preferred sessions where their teacher fully operated the system. This can be explained by the fact that the activity controlled by the teachers was also based on storytelling. For students who had previously experienced such activities, the approach was not novel, whereas students encountering it for the first time were amused by its features. This observation further reinforces the hypothesis regarding the significance of storytelling as a scenario-based learning approach.

Most of the students were able to identify the embedded elements of integration in the activities (RQ3), as their drawings and attitudes suggested. Although a deeper analysis is required on the topic, the observations reported shed light on the importance of considering inclusion elements in all educational activity designs.

The teachers' feedback and opinions were extremely positive and demonstrated the potential of our proposal to be considered as a regular strategy, answering our RQ4. The fact that the system was able to perform accurately and without taking extra time for teachers was pointed as crucial for their approval of our proposal. This serves as a powerful example of how LLMs can effectively and ethically support educational initiatives. Furthermore, teachers agreed on the advantages of using multiple models that perform better in different tasks and appreciated the fact that they increased their knowledge about LLMs.


\section{Conclusion}\label{sec:conclusion}

In this paper, we addressed a 1-week classroom validation of our system for converting regular curriculum content to scenario-based activities, with embedded integration aspects, applied to social robots. 
%
%
All the pedagogical and integrational elements we designed through prompts in our interventions, such as peer integration, individual learning, and emotional aspects, were noticed and reported by multiple children in their drawings without being directly prompted.
%

Potential pitfalls of LLMs were mitigated through an interface validation process that required human approval before being sent to the robot. This approach allowed screening of content that might approach sensitive contexts or contain potentially inaccurate information. Notably, no instances of misinformation or hallucination were observed across any of the models used.
Although a more explorative research was performed in this study, the outcomes strongly suggest the potential of the used methods to be consolidated and further investigated.
As limitations, approximating the interventions to real-world scenario did not allow for a clean collection and analyses of learning gains and children's perceptions. 
Nonetheless, the implemented architecture affords easy applications of such studies in the future.

\section*{Acknowlegdments}

We would like to thank all the teachers and children who participated in this project for their valuable time and kind cooperation. Furthermore, we would like to acknowledge the use of generative models during the preparation of this manuscript: ChatGPT 4.0 and Claude were used to identify minor grammar mistakes, enhance the readability of some paragraphs, and summarize teachers' interview transcripts. The original inputs are kept for consultation upon request.





\bibliography{references}
\bibliographystyle{ieeetr}

\end{document}